\documentclass[aps,showpacs,nofootinbib,superscriptaddress]{revtex4}
\usepackage{epsf}
\usepackage{graphicx}
\usepackage{amsmath}
\def\slashchar#1{\setbox0=\hbox{$#1$}
   \dimen0=\wd0 \setbox1=\hbox{/} \dimen1=\wd1
   \ifdim\dimen0>\dimen1 \rlap{\hbox to \dimen0{\hfil/\hfil}} #1
   \else  \rlap{\hbox to \dimen1{\hfil$#1$\hfil}} / \fi}

\begin{document}

\title{ Hyperfine mixing in $b\to c$ semileptonic decay of doubly
heavy baryons }

\author{ C.Albertus\,\footnote{Present address: Departamento de F\'\i
sica Fundamental, Universidad de Salamanca, E-37008 Salamanca,
Spain}}\affiliation{Instituto de F\'\i sica Corpuscular (IFIC), Centro
Mixto CSIC-Universidad de Valencia, Institutos de Investigaci\'on de
Paterna, Aptd. 22085, E-46071 Valencia, Spain} \author{
E. Hern\'andez} \affiliation{Departamento de F\'\i sica Fundamental e
IUFFyM,\\ Universidad de Salamanca, E-37008 Salamanca, Spain}
\author{J.~Nieves} \affiliation{Instituto de F\'\i sica Corpuscular
(IFIC), Centro Mixto CSIC-Universidad de Valencia, Institutos de
Investigaci\'on de Paterna, Aptd. 22085, E-46071 Valencia, Spain}

\pacs{12.39.Jh,13.30.Ce, 14.20.Lq, 14.20.Mr}

\begin{abstract}
  We qualitatively corroborate the results of W. Roberts and M. Pervin
 in Int. J. Mod. Phys. A 24, 2401 (2009) according to which hyperfine
 mixing greatly affects the decay widths of $b\to c$ semileptonic
 decays involving doubly heavy $bc$ baryons. However, our predictions
 for the decay widths of the unmixed states differ from those reported
 in the work of Roberts and Pervin by a factor of 2, and this
 discrepancy translates to the mixed case. We further show that 
  the predictions of heavy quark spin symmetry,
 might be used in the future to experimentally extract information on
 the admixtures in the actual physical $bc$ baryons, in a model
 independent manner.
\end{abstract}

\maketitle
%
%
%
%
%
%
\section{Introduction}
According to heavy quark spin symmetry (HQSS)~\cite{Jenkins:1992nb},
in the infinite heavy quark mass limit, one can select the heavy quark
subsystem of a doubly heavy baryon to have a well defined total spin
$S_h=0,1$. In Table~\ref{tab:dhb} we show the ground state
$J^\pi=\frac12^+,\frac32^+$ doubly heavy baryons classified so that
$S_h$ is well defined, and to which we shall refer to as the
$S_h$-basis. Being ground states for the given quantum numbers a total
orbital angular momentum $L=0$ is naturally assumed.

 Due to the finite value of the heavy quark masses, the hyperfine
interaction between the light quark and any of the heavy quarks can
admix both $S_h=0$ and $S_h=1$ spin components into the wave
function. This mixing should be negligible for $bb$ and $cc$ doubly
heavy baryons as the antisymmetry of the wave function would require
radial excitations and/or higher orbital angular momentum in the
$S_h=0$ component. On the other hand, in the $bc$ sector one expects
the actual physical $\Xi$ $(\Omega)$ particles to be admixtures of the
$\Xi_{bc},\,\Xi'_{bc}$ ($\Omega_{bc},\,\Omega'_{bc}$) states listed in
Table~\ref{tab:dhb}.  

One can minimize the effect of hyperfine mixing for $bc$ baryons by
working in a different basis, that we shall call the $qc$-basis, in
which it is the light quark $q$ (q=u,\,d,\,s) and the $c$ quark that
couple to well defined total spin $S_{qc}=0,1$. For further use we
shall denote the states in that basis as
$\hat\Xi_{bc},\,\hat\Omega_{bc}$ for $S_{qc}=1$, and
$\hat\Xi'_{bc},\,\hat\Omega'_{bc}$ for $S_{qc}=0$. In this latter
basis hyperfine mixing is always inversely proportional to the $b$
quark mass and it is thus negligible. This in fact means, that the
hyperfine mixed $bc$ states one would obtain after diagonalizing the
mass Hamiltonian in the $S_h$-basis should be very close to these new
$qc$-basis states.  The $qc$-basis was used in the doubly heavy baryon
mass determination of Ref.~\cite{roncaglia95}. However, masses are
very insensitive to hyperfine mixing and most mass
calculations~\cite{korner94,silvestre96,ebert97,itoh00,gershtein00,tong00,
mathur02, ebert02,kiselev02,narodetskii02,vijande04,albertusdhb} just
ignore the mixing and use the $S_h$-basis.

 Roberts and Pervin~\cite{pervin1} have taken up the hyperfine mixing
 issue again, pointing out that it could greatly affect the decay
 widths of doubly heavy baryons. In Ref.~\cite{elihqss}, it was later
 noticed that the $b\to c$ semileptonic decay width for transitions
 involving the $S_h$-basis $\Xi_{bc}$ ($\Omega_{bc}$) state was so
 different from the corresponding one involving the $\Xi'_{bc}$
 ($\Omega'_{bc}$) state, that experimental data, when available, could
 be used to extract information on the admixtures in the actual
 physical states. Following their own suggestion in
 Ref.~\cite{pervin1}, Roberts and Pervin have conducted a calculation
 in which they find that hyperfine mixing in the $bc$ states has a
 tremendous impact on doubly heavy baryon $b\to c$ semileptonic decay
 widths~\cite{pervin2}.  In fact this kind of information was
 partially available in the literature. In Ref.~\cite{faessler01},
 Faessler et al. evaluated the
 $\Gamma(\hat\Xi_{bc}\to\Xi_{cc}\,l\bar\nu_l)$ decay width obtaining a
 result that was a factor of around three smaller than the values obtained in
 Refs.~\cite{elihqss,ebert04} for the same transition but now
 evaluated for the $S_h$-basis $\Xi_{bc}$ state. This hinted  to the
 relevance of hyperfine mixing, but it went largely unnoticed among
 the large differences present between the different theoretical
 predictions~\cite{elihqss,ebert04,guo98,sanchis95,onishchenko00}. To
 our knowledge, Roberts and Pervin~\cite{pervin2} have been the first
 ones to realize the importance of hyperfine mixing for decay
 properties. However, their work is not sufficiently known. Thus, for
 instance, the most recent calculation of the semileptonic decay widths
 of these baryons still ignores the effects of
 mixing~\cite{faessler09}.

In the present calculation, in which we use our non-relativistic
approach described in Ref.~\cite{albertusdhb}, we try to confirm their
results.  As shown below, we qualitatively corroborate their findings
as to the importance of hyperfine mixing for evaluating $b\to c$
semileptonic decay widths of doubly heavy baryons. On the other hand,
we find large discrepancies between the two calculations for the
actual decay width values. Their results for unmixed states in the
$S_h$-basis are a factor of two smaller than ours. This discrepancy
translates to the full mixed state calculation. Besides, we will also
show how  HQSS predictions for the $b\to c$ transition matrix elements might be used in
the future to experimentally obtain information on the admixtures of  the $bc$ baryons,
in a model independent manner.
 \begin{table}
\begin{tabular}{cccc||cccc}\hline
\hspace{.25cm}Baryon\hspace{.25cm} & Quark content & \hspace{.25cm}$S_h^\pi$\hspace{.25cm} & 
\hspace{.25cm}$J^\pi$\hspace{.25cm} 
&\hspace{.25cm}Baryon\hspace{.25cm} & Quark content & \hspace{.25cm}$S_h^\pi$\hspace{.25cm} &
\hspace{.25cm}$J^\pi$\hspace{.25cm}\vspace{-.1cm}\\ 
&(l=u,d)&&&&&&\\ \hline
$\Xi_{cc}$ & \{c~c\}~l & 1$^+$ & 1/2$^+$ &\hspace{.25cm}$\Omega_{cc}$ & \{c~c\}~s & 1$^+$ & 1/2$^+$ \\ 
$\Xi_{cc}^*$ & \{c~c\}~l & 1$^+$ & 3/2$^+$&\hspace{.25cm}$\Omega_{cc}^*$ & \{c~c\}~s & 1$^+$ & 3/2$^+$ \\ 
$\Xi_{bb}$ & \{b~b\}~l & 1$^+$ & 1/2$^+$ &\hspace{.25cm}$\Omega_{bb}$ & \{b~b\}~s & 1$^+$ & 1/2$^+$ \\ 
$\Xi_{bb}^*$ & \{b~b\}~l & 1$^+$ & 3/2$^+$&\hspace{.25cm}$\Omega_{bb}^*$ & \{b~b\}~s & 1$^+$ & 3/2$^+$ \\ 
$\Xi_{bc}$ & \{b~c\}~l & 1$^+$ & 1/2$^+$ &\hspace{.25cm}$\Omega_{bc}$ & \{b~c\}~s & 1$^+$ & 1/2$^+$ \\ 
$\Xi_{bc}^*$ & \{b~c\}~l & 1$^+$ & 3/2$^+$&\hspace{.25cm}$\Omega_{bc}^*$ & \{b~c\}~s & 1$^+$ & 3/2$^+$ \\ 
$\Xi_{bc}'$ & [b~c]~l & 0$^+$ & 1/2$^+$ &\hspace{.25cm}$\Omega_{bc}'$ & [b~c]~s & 0$^+$
& 1/2$^+$ \\ \hline
\end{tabular}%
\caption{Quantum numbers and quark content for ground state doubly heavy baryons classified so that $S_h$
(spin of the heavy quark subsystem) is
well defined. }
\label{tab:dhb}
\end{table}

%
%
%
%
%
%
%
%
%
\section{Results and discussion}
As in Refs.~\cite{albertusdhb,elihqss}, we shall work in the
 $S_h$-basis. We assume $L=0$ and symmetric orbital wave functions
 throughout, neglecting thus hyperfine mixing for doubly $bb$ and $cc$
 baryons. However we shall take into account hyperfine mixing in the
 $bc$ sector.  Our unmixed wave functions are obtained using a
 variational approach that is based on HQSS. A full account can be
 found in Ref.~\cite{albertusdhb}.  We shall use the AL1 potential of
 B. Silvestre-Brac and C. Semay~\cite{semay94,silvestre96} that for
 the $q\bar q $ interaction reads
\begin{eqnarray}
V_{q\bar q}(r) = -\frac{\kappa}{r}
+\lambda r - \Lambda + \frac{2\pi}{3m_q m_{\bar q}}\kappa^\prime  \frac{e^{-r^2/x_0^2}}{\pi^\frac32 x_0^3} 
\ \vec{\sigma}_q\cdot\vec{\sigma}_{\bar q} \ \ ; \ \ 
x_0(m_q,m_{\bar q}) = A \left ( \frac{2m_qm_{\bar q}}{m_q+m_{\bar q}} \right )^{-B}.
\label{eq:al1}
\end{eqnarray}
It contains a linear confinement term, plus Coulomb and hyperfine
($\vec\sigma\cdot\vec\sigma$) terms coming from one-gluon exchange.
All free parameters in the potential had been adjusted to reproduce
the light and heavy-light meson spectra. For its use in baryons we
apply the usual prescription~\cite{silvestre96,badhuri81}
\begin{equation}
V_{qq}=\frac12\,V_{q\bar q}
\end{equation}
This is fully justified for the one-gluon exchange part of the
potential where the implicit color dependence accounts for the factor
of two difference between the $q\bar q$ and the $qq$ interaction. The
prescription is also phenomenologically very successfully for the
confinement part. As shown in the lattice QCD calculation of
Ref.~\cite{suganuma05}, the confinement part of the potential for a
three quark system is proportional to the minimal total length $
L_{\rm min}$ of the color flux tube linking the three quarks, and in
fact one has $\frac12\sum_{j<k}|\vec r_j-\vec r_k|\approx { L}_{\rm
min}$. The theoretical uncertainties on masses and decay widths
associated to the use of different potentials were found to be small
in Ref.~\cite{albertusdhb}, where we presented results for five
different potentials taken from
Refs.~\cite{semay94,silvestre96,badhuri81}\footnote{All five
potentials are conveniently tabulated in Ref.~\cite{conrado04}. There,
and for the first time, a variational approach based on HQSS was
applied to solve the three-body problem in heavy baryons.  A similar
approach was used in Ref.~\cite{albertusdhb} for doubly heavy
baryons.}.

To evaluate the decay widths we shall use the model described in
Ref.~\cite{albertusdhb}. We work in a spectator approximation in which
any of the $b$ quarks in the initial state can decay into any of the
$c$ quarks in the final state. This, together with the right symmetry
factor for doubly heavy (initial or final) baryon states with two
equal heavy quarks, gives rise to an extra factor $\sqrt2$ in the
amplitude compared to the similar $b\to c$ decays for baryons with
just one heavy quark. This is in fact more important than it seems as
many theoretical calculations have overlooked that factor, or got it
wrong, altogether (See the erratum in Ref.~\cite{albertusdhb}).
%
%
%
%
%
%
%
\subsection{Masses: unmixed results}
In Table~\ref{tab:munmixed} we present the results we obtain for the
masses of $S_h$-basis unmixed states. We have corrected numerical
inaccuracies present in our former work in Ref.~\cite{albertusdhb} and
the masses we give now deviate slightly from the ones reported
there. As a result, small changes will affect  the decay widths to
be discussed below. We also show the results obtained by Roberts and
Pervin in Ref.~\cite{pervin1}. They use a non-relativistic approach in
which the orbital wave functions are expanded in a harmonic oscillator
basis. Their results are always larger than ours by $50\sim
180\,$MeV. For the sake of comparison we also show the results of the
non-relativistic calculation in Ref.~\cite{gershtein00}, that give
results that are about 100\,MeV smaller that ours, and the ones obtained
in three different relativistic approaches~\cite{ebert02,he04,martynenko08}.
\begin{table}
\begin{tabular}{lcccccc||lcccccc}\hline
& This work&\cite{pervin1}&\cite{gershtein00}&\cite{ebert02} &\cite{he04}&\cite{martynenko08}\hspace{.25cm}
& &This work&\cite{pervin1}&\cite{gershtein00}&\cite{ebert02}&\cite{he04}&\cite{martynenko08}\\ \hline
$M_{\Xi_{cc}}$      &3613 & 3676&3478&3620&3550&3510 \hspace{.25cm}    
& \hspace{.25cm}$M_{\Omega_{cc}}$        & 3712 &3815&3590&3778 &3730&3719\\
$M_{\Xi_{cc}^*}$    &3707 &3753&3610&3727&3590&3548  \hspace{.25cm}    
& \hspace{.25cm}$M_{\Omega_{cc}^*}$        & 3795 &3876 &3690& 3872&3770&3746 \\ 
$M_{ \Xi_{bb}} $    &10198 &10340&10093 &10202&10100&10130 \hspace{.25cm} 
& \hspace{.25cm}$M_{\Omega_{bb}}$         & 10269 &10454&10180&10359&10280&10422 \\ 
$M_{\Xi_{bb}^*}$     &10237 &10367&10133 &10237&10110&10144 \hspace{.25cm}
& \hspace{.25cm}$M_{\Omega_{bb}^*}$        & 10307 &10486&10200&10389&10290&10432\\ 
$M_{\Xi_{bc}}$       &6928 &7020 &6820&6933&6800&6792$^\dagger$ \hspace{.25cm}   
& \hspace{.25cm}$M_{\Omega_{bc}}$         & 7013  &7147&6910&7088&6980&6999$^\dagger$ \\ 
$M_{\Xi_{bc}'}$      &6958 & 7044&6850 & 6963&6870&6825$^\dagger$  \hspace{.25cm}
&\hspace{.25cm}$M_{\Omega_{bc}'}$           & 7038 &7166&6930&7116&7050&7022$^\dagger$\\
$M_{\Xi_{bc}^*}$   &6996  &7078&6900 &6980&6850&6827   \hspace{.25cm}  
&\hspace{.25cm}$M_{\Omega_{bc}^*}$        & 7075 &7191&6990&7130&7020&7024\\
\hline
\end{tabular}%
\caption{Masses (MeV units) for $S_h$-basis unmixed states.  Mixing has
been taken into account only for the results marked with a $^\dagger$ symbol.}
\label{tab:munmixed}
\end{table}
On the experimental side the SELEX Collaboration claimed evidence for
the $\Xi^+_{cc}$ baryon, in the $\Lambda_c^+K^-\pi^+$ and $pD^+K^-$
decay modes, with a mass of $M_{\Xi^+_{cc}}=3519\pm 1\
\mathrm{MeV/c^2}$~\cite{mattson}. No other experimental collaboration
has found evidence for doubly charmed baryons so far and, at present,
the $\Xi^+_{cc}$ has only a one star status.%
%
%
%
%
%
\subsection{Semileptonic $b\to c$ decays: unmixed results}
In Table~\ref{tab:gunmixed} we show the results for the $b\to c$
semileptonic decay widths of doubly heavy baryons for unmixed states
in the $S_h$-basis, and assuming massless fermions in the final state,
thus only valid for decays into $e,\mu$ but not into $\tau$. Our
results are roughly a factor of two larger than the ones obtained by
Roberts and Pervin in Ref.~\cite{pervin2}. This discrepancy can not be
attributed to the small mass differences between the two calculations
as the energy involved in $b\to c$ decays is very large. We also show
the decay widths obtained in two different relativistic
approaches~\cite{ebert04,faessler09}.  Our results are in a global
fair agreement with the ones in Ref.~\cite{ebert04}. The agreement is
also good with Ref.~\cite{faessler09} but only for transitions with a
$bc$ baryon in the initial state.  For transitions with a $bb$ baryon
in the initial state, hence a $bc$ baryon in the final state, their
results are a factor of two smaller than ours.  
 All four calculations comply with the HQSS constraints
among decay width ratios derived in Ref.~\cite{elihqss} for unmixed
states in the $S_h$-basis. As only ratios are involved, those
relations can not be used to elucidate which of the non-relativistic
calculation is preferable. Besides, as the Isgur-Wise functions are
different, the ratios concern separately $bb\to bc$ decays and $bc\to
cc$ decays so they can not be used to see which relativistic
calculation, if any, is more correct for the case of the $bb\to bc$
transitions. This constitutes  an open problem.
\begin{table}
\begin{tabular}{lcccc||lcccc}\hline
& This
work&\hspace*{.25cm}\cite{pervin2}\hspace*{.25cm}&\hspace*{.25cm}\cite{ebert04}\hspace*{.25cm}&\cite{faessler09}\hspace{.25cm}
&&This
work&\hspace*{.25cm}\cite{pervin2}\hspace*{.25cm}&\hspace*{.25cm}\cite{ebert04}\hspace*{.25cm}&\cite{faessler09}\\\hline
$\Gamma(\Xi_{bb}^*\to\Xi_{bc}'\,l\bar\nu_l)$ &  $1.08$ &-- &$0.82$&
$0.36\pm0.10$\hspace{.25cm}
&\hspace{.25cm}$\Gamma(\Omega_{bb}^*\to\Omega_{bc}'\,l\bar\nu_l)$ & $1.14$&--
&$0.85$&$0.42\pm0.14$ \\ 
$\Gamma(\Xi_{bb}^*\to\Xi_{bc}\,l\bar\nu_l)$
&$0.36$&--&$0.28$&$0.14\pm0.04$\hspace{.25cm}
&\hspace{.25cm}$\Gamma(\Omega_{bb}^*\to\Omega_{bc}\,l\bar\nu_l)$
&$0.38$&--&$0.29$&$0.15\pm0.05$\\ 
$\Gamma(\Xi_{bb}\to\Xi_{bc}'\,l\bar\nu_l)$ &  $1.09$ &$0.41$&$0.82$&$0.43\pm0.12$\hspace{.25cm}

&\hspace{.25cm}$\Gamma(\Omega_{bb}\to\Omega_{bc}'\,l\bar\nu_l)$ & $1.16$&$0.51$
&$0.83$&$0.48\pm0.12$ \\ 
$\Gamma(\Xi_{bb}\to\Xi_{bc}\,l\bar\nu_l)$  &$2.00$&$0.69$&$1.63$&$0.80\pm0.30$\hspace{.25cm}
&\hspace{.25cm}$\Gamma(\Omega_{bb}\to\Omega_{bc}\,l\bar\nu_l)$  &
$2.15$&$0.92$&$1.70$&$0.86\pm0.32$\\ 
$\Gamma(\Xi_{bc}'\to\Xi_{cc}\,l\bar\nu_l)$ & 
$1.36$&--&$0.88$&$1.10\pm0.32$\hspace{.25cm}
&\hspace{.25cm}$\Gamma(\Omega_{bc}'\to\Omega_{cc}\,l\bar\nu_l)$ &
$1.36$&--&$0.95$&$0.98\pm0.28$ \\ 
$\Gamma(\Xi_{bc}\to\Xi_{cc}\,l\bar\nu_l)$  &$ 2.57
$&$1.38$&$2.30$&$2.10\pm0.70$\hspace{.25cm}
&\hspace{.25cm}$\Gamma(\Omega_{bc}\to\Omega_{cc}\,l\bar\nu_l)$  &
$2.58$&$1.54$&$2.48$&$1.88\pm0.62$\\ 
$\Gamma(\Xi_{bc}'\to\Xi_{cc}^*\,l\bar\nu_l)$ & 
$2.35$&--&$1.70$&$2.01\pm0.62$ \hspace{.25cm}
&\hspace{.25cm}$\Gamma(\Omega_{bc}'\to\Omega_{cc}^*\,l\bar\nu_l)$ & $2.35$
&--&$1.83$&$1.93\pm0.60$\\ 
$\Gamma(\Xi_{bc}\to\Xi_{cc}^*\,l\bar\nu_l)$  &$ 0.75 $&$0.52$ &$0.72$&$0.64\pm0.19$\hspace{.25cm}
&\hspace{.25cm}$\Gamma(\Omega_{bc}\to\Omega_{cc}^*\,l\bar\nu_l)$  &$0.76$&$0.56$&$0.74$&$0.62\pm0.19$\\\hline
\end{tabular}
\caption{ Semileptonic decay widths (in units of $10^{-14}\ {\rm
 GeV}$) for $S_h$-basis unmixed states.  We have used
 $|V_{cb}|=0.0413$. $l$ stands for $l=e,\mu$. }
\label{tab:gunmixed}%
\end{table}

An interesting feature of the decay widths shown in
Table~\ref{tab:gunmixed} is that they are very different for
transitions involving $\Xi_{bc}$ or $\Xi'_{bc}$ (idem $\Omega _{bc}$
or $\Omega'_{bc}$). Thus, one expects that the decay widths involving
the actual physical $bc$ states could be very different from the ones
quoted in Table~\ref{tab:gunmixed}, provided hyperfine mixing in the
$bc$ baryons is non negligible. In fact, we anticipate this to be the
case as for $bc$ states one expects physical $\Xi$ and $\Omega$ to be
close to the $qc$-basis states for which, apart from a global phase,
one has (in what follows $B\equiv\Xi$ or $\Omega$)
\begin{eqnarray}
\hat B_{bc}&=&\frac{\sqrt3}{2}B'_{bc}+\frac{1}{2}B_{bc},\nonumber\\
\hat B'_{bc}&=&-\frac{1}{2}B'_{bc}+\frac{\sqrt3}{2}B_{bc}.
\label{eq:qchqss}
\end{eqnarray}%
%
%
%
%
%
%
%
\subsection{Results with mixing}
To obtain the physical $bc$ states in the $S_h$-basis we have to
diagonalize the mass matrices for which the diagonal matrix elements
are the corresponding masses in Table \ref{tab:munmixed}. The
hyperfine $\vec\sigma\cdot\vec\sigma$ term (see Eq.(\ref{eq:al1})) in
the interaction between the light and any of the heavy quarks is
responsible for the mixing as it gives rise to non vanishing non
diagonal matrix elements. The values for the latter are respectively
$18.3\,$MeV for $\Xi$ baryons and $15.8\,$MeV for $\Omega$
baryons. After diagonalizing we get for the physical
$\Xi_{bc}^{(1)}$ and $\Xi_{bc}^{(2)}$ states
\begin{eqnarray}
\Xi_{bc}^{(1)}&=&\hspace{.25cm}0.902\ \Xi'_{bc}+0.431\ \Xi_{bc}\ \ ;\ M_{\Xi_{bc}^{(1)}}=6967\,{\rm MeV},\nonumber\\
\Xi_{bc}^{(2)}&=&-0.431\  \Xi'_{bc}+0.902\ \Xi_{bc}\ \ ;\ M_{\Xi_{bc}^{(2)}}=6919\,{\rm MeV},
\label{eq:mixedxi}
\end{eqnarray}
for $\Xi$ baryons, and
\begin{eqnarray}
\Omega_{bc}^{(1)}&=&\hspace{.25cm}0.899\  \Omega'_{bc}+0.437\ \Omega_{bc}\ \ ;\ M_{\Omega_{bc}^{(1)}}=7046\,{\rm MeV},\nonumber\\
\Omega_{bc}^{(2)}&=&-0.437\  \Omega'_{bc}+0.899\ \Omega_{bc}\ \ ;\ M_{\Omega_{bc}^{(2)}}=7005\,{\rm MeV},
\label{eq:mixedomega}
\end{eqnarray} 
for $\Omega$ baryons. Theoretical uncertainties related to the
determination of the variational wave functions may affect the last of
the digits quoted above. As one sees, the masses of the physical
states change only by a few MeV
compared to the corresponding unmixed values in
Table~\ref{tab:munmixed}. On the other hand the mixture is important
and, as mentioned before, physical states are close to the $qc$-basis
states (see Eq.(\ref{eq:qchqss})).

 We have evaluated again the decay widths for the mixed states. These
 results are displayed in Table~\ref{tab:gmixed} where, for better
 comparison, we also show in parenthesis the corresponding unmixed
 results. The changes we find in decay widths are very large,
 confirming the results of Roberts and Pervin in
 Ref.~\cite{pervin2}. Qualitatively we find the same behavior, but the
 actual decay widths are very different. This is a reflection of the
 factor of two discrepancy present already for the unmixed case.
 
 As noticed in Ref.~\cite{pervin2}, the widths for the decays
 $B^{(2)}_{bc}\to B^*_{cc}$ decrease considerably from their unmixed
 counterparts. In our calculation the decrease factor is $44$ (54) for
 the $\Xi^{(2)}_{bc}\to\Xi^*_{cc}$ ($\Omega^{(2)}_{bc}\to
 \Omega^*_{cc}$) transition.  This can be easily understood by taking
 into account that $B^{(2)}_{bc}\approx \hat B'_{bc}$. In the latter
 state, the light and $c$ quarks are coupled to spin 0, whereas in the
 $B^*_{cc}$ the light and any of the $c$ quarks are in a relative spin
 1 state. In any spectator calculation, as the ones here and in
 Ref.~\cite{pervin2}, the amplitude for the $\hat B'_{bc}\to B^*_{cc}$
 transition cancels due to the orthogonality of the different spin
 states of the spectator quarks ($qc$ pair) in the initial and final
 baryons. The fact that $B^{(2)}_{bc}$ slightly deviates from $ \hat
 B'_{bc}$ produces a non zero, but small, decay width.
%
%
%
%
%
%
\subsection{HQSS and mixing}
  The last result 
 can be derived on more general grounds in the context of HQSS. In
 Ref.~\cite{flynn07}, using HQSS, the following hadronic amplitudes
 were found for states in the $S_h$-basis
\begin{eqnarray}
\left.{\cal A}(B_{bc}\to B^*_{cc})\right|_{\rm HQSS}&=&\frac{1}{\sqrt2}\,\frac{-2}{\sqrt3}\eta\ \bar
u'^\mu \ u,\nonumber\\
\label{eq:hqss4}
\left.{\cal A}(B_{bc}'\to B^*_{cc})\right|_{\rm HQSS}&=&\frac{1}{\sqrt2}\,(-2)\,\eta\ \bar
u'^\mu \ u,
\end{eqnarray}
where $\eta$ is the universal Isgur-Wise function, different for $\Xi$
and $\Omega$ decays, and $u,u^\mu$ are respectively the Dirac and
Rarita-Schwinger spinors for the initial and final baryons. Combining
these results together with Eq.(\ref{eq:qchqss}) one immediately
derives
\begin{eqnarray}
\left.{\cal A}(\hat B_{bc}'\to B^*_{cc})\right|_{\rm HQSS}=0 \label{eq:eqj1}
\end{eqnarray}
implying $\Gamma(\hat B'_{bc}\to B^*_{cc})=0$  as before.
\begin{table}
\begin{tabular}{lcccc||lcccc}\hline
 &This work &This work&\cite{pervin2}&\cite{pervin2}\hspace{.2cm}&
 &This work&This work&\cite{pervin2}&\cite{pervin2}\\
  &mixed&unmixed&mixed&unmixed&&mixed&unmixed&mixed&unmixed\\\hline
$\Gamma(\Xi_{bb}^*\hspace{.101cm} \to\Xi^{(1)}_{bc}\,l\bar\nu_l)$ &  $0.47$&(1.08)& -- & --\hspace{.2cm} 
&\hspace{.2cm}$\Gamma(\Omega_{bb}^*\hspace{.1cm}\to\Omega^{(1)}_{bc}\,l\bar\nu_l)$ & $0.48$&(1.14)&-- &-- \\ 
$\Gamma(\Xi_{bb}^*\hspace{.101cm}\to\Xi^{(2)}_{bc}\,l\bar\nu_l)$\ &$0.99$&(0.36)&--&--\hspace{.2cm}
&\hspace{.2cm}$\Gamma(\Omega_{bb}^*\hspace{.1cm}\to\Omega^{(2)}_{bc}\,l\bar\nu_l)$\ &$1.06$&(0.38)&--&--\\ 
$\Gamma(\Xi_{bb}\hspace{.101cm}\to\Xi^{(1)}_{bc}\,l\bar\nu_l)$ &  $2.21$&(1.09)& $0.95$&(0.41) \hspace{.2cm}
&\hspace{.2cm}$\Gamma(\Omega_{bb}\hspace{.1cm}\to\Omega^{(1)}_{bc}\,l\bar\nu_l)$ & $2.36$&(1.16)&
$0.99$&(0.51)  \\ 
$\Gamma(\Xi_{bb}\hspace{.101cm}\to\Xi^{(2)}_{bc}\,l\bar\nu_l)$  &$0.85$&(2.00)& $0.33$&(0.69)\hspace{.2cm}
&\hspace{.2cm}$\Gamma(\Omega_{bb}\hspace{.1cm}\to\Omega^{(2)}_{bc}\,l\bar\nu_l)$  & $0.91$&(2.15)& $0.30$&(0.92)\\ 
$\Gamma(\Xi^{(1)}_{bc}\to\Xi_{cc}\hspace{.101cm}\,l\bar\nu_l)$ &  $0.38$&(1.36)& --& --\hspace{.2cm}
&\hspace{.2cm}$\Gamma(\Omega^{(1)}_{bc}\to\Omega_{cc}\hspace{.1cm}\,l\bar\nu_l)$ & $0.37$&(1.36)& --& -- \\ 
$\Gamma(\Xi^{(2)}_{bc}\to\Xi_{cc}\hspace{.101cm}\,l\bar\nu_l)$  &$ 3.50$&(2.57)&$ 1.92 $&(1.38)\hspace{.2cm}
&\hspace{.2cm}$\Gamma(\Omega^{(2)}_{bc}\to\Omega_{cc}\hspace{.1cm}\,l\bar\nu_l)$  & $3.52$&(2.58)& $1.99$&(1.54)\\ 
$\Gamma(\Xi^{(1)}_{bc}\to\Xi_{cc}^*\hspace{.101cm}\,l\bar\nu_l)$ &  $3.14$&(2.35)&--&--\hspace{.2cm}
&\hspace{.2cm}$\Gamma(\Omega^{(1)}_{bc}\to\Omega_{cc}^*\hspace{.1cm}\,l\bar\nu_l)$ & $3.14$&(2.35)&--&-- \\ 
$\Gamma(\Xi^{(2)}_{bc}\to\Xi_{cc}^*\hspace{.101cm}\,l\bar\nu_l)$  &$0.017$&(0.75)& $0.026$&(0.52)\hspace{.2cm}
&\hspace{.2cm}$\Gamma(\Omega^{(2)}_{bc}\to\Omega_{cc}^*\hspace{.1cm}\,l\bar\nu_l)$ 
&$0.014$&(0.76)&$0.013$&(0.56)\\\hline
\end{tabular}
\caption{ Semileptonic decay widths (in units of $10^{-14}\ {\rm
GeV}$) for mixed states. For better comparison we also show in
parenthesis the results for unmixed states. $l=e,\mu$.  We have used
$|V_{cb}|=0.0413$.}
\label{tab:gmixed}
\end{table}%
For completeness, we give all the hadronic amplitudes for decays
involving $\hat B_{bc},\hat B'_{bc}$ that derive from the HQSS
relations in Ref.~\cite{flynn07}
\begin{eqnarray}
\left.{\cal A}(\hat B_{bc}\to B_{cc})\right|_{\rm HQSS}&=&\frac{\eta}{\sqrt2}\,\bar
u'(\gamma^\mu+\frac13\,\gamma^\mu\gamma_5)\,u,\nonumber\\
\left.{\cal A}(\hat B'_{bc}\to B_{cc})\right|_{\rm HQSS}&=&\frac{\eta}{\sqrt2}\,\sqrt3\,\bar
u'(\gamma^\mu-\gamma^\mu\gamma_5)\,u,\nonumber\\
\left.{\cal A}(\hat B_{bc}\to B^*_{cc})\right|_{\rm HQSS}&=&\frac{\eta}{\sqrt2}\,\frac{-4}{\sqrt3}\,\bar
u'^\mu\,u.\nonumber\\ \label{eq:eqj2}
\end{eqnarray}
Similarly one has 
\begin{eqnarray}
\left.{\cal A}(B_{bb}\to \hat B_{bc})\right|_{\rm HQSS}&=&\frac{\eta'}{\sqrt2}\,\bar
u'(\gamma^\mu-\frac53\,\gamma^\mu\gamma_5)\,u,\nonumber\\
\left.{\cal A}( B_{bb}\to\hat  B'_{bc})\right|_{\rm HQSS}&=&\frac{\eta'}{\sqrt2}\,\sqrt3\,\bar
u'(\gamma^\mu-\frac13\gamma^\mu\gamma_5)\,u,\nonumber\\
\left.{\cal A}(B^*_{bb}\to \hat B_{bc})\right|_{\rm HQSS}&=&  \frac{\sqrt2}{\sqrt3}\,\eta'\,\bar u'\,u^\mu,\nonumber\\
\left.{\cal A}( B^*_{bb}\to\hat  B'_{bc})\right|_{\rm
  HQSS}&=&-\sqrt2\,\eta'\,\bar u'\,u^\mu.\nonumber\\ \label{eq:eqjn3}
\end{eqnarray}
where $\eta'$ is different from $\eta$ and, as before, we have a different $\eta'$ for $\Xi$ and $\Omega$
decays. 

Following the steps and approximations in Ref.~\cite{elihqss} one can derive HQSS-based relations for ratios  
involving decay widths of $qc$-basis states. Thus, for instance
\begin{eqnarray}
\left.R_1=\frac{\Gamma(\hat B'_{bc}\to  B_{cc})}
{\frac{3}{4}\Gamma( \hat B_{bc}\to  B^*_{cc})+3\Gamma(\hat B_{bc}\to  B_{cc})}\right|_{\rm HQSS}
=1,
\label{eq:ratio1}
\end{eqnarray}
and
\begin{eqnarray}
&&\hspace{1.5cm}\left.R_2=\frac{3\Gamma( B^*_{bb}\to\hat  B_{bc})}{\Gamma( B^*_{bb}\to\hat  B'_{bc})}\right|_{\rm HQSS}=1,\nonumber\\
&&\left.R_3=\frac{\frac{56}{9}\Gamma( B^*_{bb}\to\hat  B_{bc})+\frac{10}{27}\Gamma( B_{bb}\to\hat  B'_{bc})}
{\Gamma( B_{bb}\to\hat  B_{bc})}\right|_{\rm HQSS}=1.
\label{eq:ratio2}
\end{eqnarray}

We can evaluate those ratios for the actual physical states
substituting $\hat B_{bc}$ by $B_{bc}^{(1)}$ and $\hat B'_{bc}$ by
$B_{bc}^{(2)}$ in Eqs.(\ref{eq:ratio1}) and (\ref{eq:ratio2}), and
using the widths in Table~\ref{tab:gmixed}. What one finds is
$R_1=1.00,\,R_2= 1.42$ and $R_3=1.47$. The sizeable deviation from 1
in the two latter cases is due to the difference between the physical
mixing coefficients in Eqs. (\ref{eq:mixedxi}) and
(\ref{eq:mixedomega}) and the ideal ones in Eq. (\ref{eq:qchqss}),
while the extraordinary agreement for $R_1$ seems to be purely
accidental.  To check that assertion we have played the game of
evaluating the decay widths involving $\hat B_{bc},\hat B'_{bc}$
states, assuming their masses to be respectively the ones for $
B^{(1)}_{bc}, B^{(2)}_{bc}$. What we get for the ratios in this case
is $R_1=1.07,\,R_2= 0.996$ and $R_3=1.06$, all of them in good
agreement with the HQSS-based predictions in Eqs.(\ref{eq:ratio1}) and
(\ref{eq:ratio2}). 

HQSS can predict the hadronic amplitudes of given states, for instance
the ones in Ref.~\cite{elihqss} for $S_h$-basis states, or the ones
here that involve $qc$-basis states, but it does not tell which are
the physical states. Thus in order to use HQSS to check width ratios
for physical states, those states have to be known
beforehand. 

However, HQSS predictions might be used to
experimentally obtain information on the mixing angle for $bc$ baryons in a model
independent manner.
Physical and $S_h$ or
$qc$-basis states will differ in a rotation, determined by a certain
mixing angle. For instance one could write
\begin{eqnarray}
\left (\begin{array}{c}B^{(1)}_{bc} \\ \\
  B^{(2)}_{bc}\end{array}\right) &=& \left(\begin{array}{cc}\cos\theta &
  \sin\theta  \\ \\
 -\sin\theta &
  \cos\theta\end{array}\right) \left (\begin{array}{c}\hat B_{bc} \\ \\
  \hat B'_{bc}\end{array}\right),
\end{eqnarray}
where one expects $\theta$ to be small. If  the ratios 
\begin{equation}
R_1^{\rm phys.}=\frac{\Gamma( B^{(2)}_{bc}\to B^*_{cc})}{\Gamma(
    B^{(1)}_{bc}\to B^*_{cc})}\ \ ,\ \ R_2^{\rm phys.}=\frac{\Gamma(
    B^*_{bb}\to\hat B^{(1)}_{bc})}{\Gamma( B^*_{bb}\to\hat
    B^{(2)}_{bc})},
\end{equation}
were measured, we would have from the HQSS relations in Eqs.(\ref{eq:eqj1}),
(\ref{eq:eqj2}) and (\ref{eq:eqjn3})
\begin{eqnarray}
R_1^{\rm phys.} &\sim&  (\tan\theta)^2 + {\cal
  O}(\frac{m_q,\Lambda_{QCD}}{m_c}), \nonumber\\
R_2^{\rm phys.} &\sim & \left (\frac{1 -\sqrt3 \tan\theta}{\sqrt3+\tan\theta }
\right)^2 + {\cal
  O}(\frac{m_q,\Lambda_{QCD}}{m_c}), 
\end{eqnarray}
which would allow to determine\footnote{For finite charm and bottom
masses, there would be some short distance corrections, which 
 probably largely cancel out in the ratio of widths.} the mixing
angle $\theta$ in a model independent manner, since HQSS is a proper
QCD spin--flavor symmetry when the quark masses become much larger
than the typical confinement scale, $\Lambda_{\rm QCD}$.  To show that
this idea works, we evaluate those ratios in the $\Xi$ sector with our
predictions for the decay widths. From them we get the mixing angle
$\theta$ to be $|\theta|=4.2^{\rm o}$ from $R_1^{\rm phys.}$ and
$\theta=-4.6^{\rm o}$ from $R_2^{\rm phys.}$. In this latter case, we
have taken the solution associated to a small mixing angle
($\tan\theta \le 1/\sqrt 3$). These estimates should be compared with
$\theta=-4.5^{\rm o}$  as deduced from
Eqs.(\ref{eq:qchqss}) and (\ref{eq:mixedxi}). Similar results are
obtained in the $\Omega$ sector. These ideas were firstly proposed in
Ref.~\cite{elihqss}.

\section{Conclusion}

In the limit of charm and bottom masses infinitely large, though
different, all spin--spin interactions can be neglected, and then
there would exist  a $J=1/2$ baryon $\Xi^{\infty}_{bc}$ with a total
degeneracy\footnote{The discussion is similar for the case of
  $\Omega$ baryons.} of 4. A basis in this space can be
constructed out of the states of the $S_h$-basis: $| J=1/2, M=\pm 1/2;
\Xi_{bc}\rangle$ and $| J=1/2, M=\pm 1/2; \Xi'_{bc}\rangle$. The
semileptonic $b\to c$ width of the $\Xi^{\infty}_{bc}$ baryon would
then depend of the particular state (an arbitrary linear combination of
the  elements of the basis) in which it is prepared, in the
same manner as for instance, the polarized $\Lambda \to N\pi$ decay
width depends on the polarization of the initial hyperon. Thus, the
HQSS relations found in \cite{flynn07} for the states of the
$S_h$-basis can be used to derive HQSS relations for the transitions
of any other $\Xi^{\infty}_{bc}$ baryon state (for instance
Eqs.(\ref{eq:eqj1}) and~(\ref{eq:eqj2}) for the case of states  with
well defined spin for the light--charm quark pair). Analogous
considerations apply for the decays of the $bb-$baryons
into a particular $\Xi^{\infty}_{bc}$ baryon state. 

In the real world, the charm and bottom masses, though large, are
finite and the above degeneration is reduced, and thus one is left to
that associated to third component of the total spin of the
baryon. Because of the hyperfine interactions, the $\Xi_{bc}$ and
$\Xi'_{bc}$ states are not eigenstates of the Hamiltonian, which is
non-diagonal in the $S_h$-basis. We have diagonalized the Hamiltonian
and found that the physical states are quite similar to those where
the spin of the light--charm quark pair is well defined. This is
because the charm quark mass is substantially smaller than that of the
bottom one. 

We have shown that this hyperfine mixing greatly affects the decay widths of
$b\to c$ semileptonic decays involving doubly heavy $bc$ baryons, and
thus we have qualitatively corroborated the results of W. Roberts and
M. Pervin of Ref. ~\cite{pervin2}. However, our predictions for
 the decay widths of the unmixed states differ from those reported in
 the work of Roberts and Pervin by a factor of 2, and this discrepancy
 translates to the mixed case.

 Finally, we have discussed how the HQSS predictions for the
 semileptonic decay widths might be used in the future to
 experimentally extract information on the admixtures in the actual
 physical $bc-$baryon states. 

%
%
%
%
%
%
%
\begin{acknowledgments}
  This research was supported by DGI and FEDER funds, under contracts
  FIS2008-01143/FIS, FIS2006-03438, FPA2007-65748, and the Spanish
  Consolider-Ingenio 2010 Programme CPAN (CSD2007-00042), by Junta de
  Castilla y Le\'on under contracts SA016A07 and GR12, and by the EU
  HadronPhysics2 project. C. A. wishes to acknowledge a contract 
  supported by PIE-CSIC 200850I238 during his stay at IFIC where part of this
  work was done.
\end{acknowledgments}

\end{document}